\documentclass[a4paper,fleqn]{cas-sc}

\usepackage[authoryear]{natbib}
\usepackage{lineno}
\usepackage{chemformula}
\usepackage[dvipsnames]{xcolor}
\def\tsc#1{\csdef{#1}{\textsc{\lowercase{#1}}\xspace}}
\tsc{WGM}
\tsc{QE}

\begin{document}
\let\WriteBookmarks\relax
\def\floatpagepagefraction{1}
\def\textpagefraction{.001}

\newcommand{\ex}[0]{\varepsilon_z}
\newcommand{\vs}[0]{\upsilon}
\newcommand{\fk}[0]{f_\kappa}
\newcommand{\xil}[0]{\xi_\lambda}

\newcommand{\aj} {"Astron. J."} 
\newcommand{\actaa} {"Acta Astron."}
\newcommand{\araa} {"Annu. Rev. Astron. Astrophys."}
\newcommand{\apj}  {"Astrophys. J."}
\newcommand{\apjl} {"Astrophys. J. Lett."}
\newcommand{\apjs} {"Astrophys. J. Suppl. S"}
\newcommand{\ao} {"Appl. Optics"}
\newcommand{\apss} {"Astrophys. Space Sci."}
\newcommand{\aap} {"Astronom. Astrophys."}
\newcommand{\aapr} {"Astron. Astrophys Rev"}
\newcommand{\aaps} {"Astron. Astrophys. Sup."}
\newcommand{\azh} {"Astron. Zh+"}
\newcommand{\caa} {"Chinese Astron. Astr."}
\newcommand{\icarus} {"Icarus"}
\newcommand{\jcap} {"J. Cosmol. Astropart. Phys."}
\newcommand{\jrasc} {"J. Roy. Astron. Soc. Can."}
\newcommand{\memras} {"Memoirs of the RAS"}
\newcommand{\mnras} {"Mon. Not. R. Astron. Soc."}
\newcommand{\na} {"New Astron."}
\newcommand{\nar} {"New Astron. Rev."}
\newcommand{\pra} {"Phys. Rev. A"}
\newcommand{\prb} {"Phys. Rev. B"}
\newcommand{\prc} {"Phys. Rev. C"}
\newcommand{\prd} {"Phys. Rev. D"}
\newcommand{\pre} {"Phys. Rev. E"}
\newcommand{\prl} {"Phys. Rev. Lett"}
\newcommand{\pasa} {"Publ. Astron. Soc. Aust."}
\newcommand{\pasp} {"Publ. Astron. Soc. Pac."}
\newcommand{\pasj} {"Publ. Astron. Soc. Jpn."}
\newcommand{\rmxaa} {"Rev. Mex. Astron. Astr."}
\newcommand{\rjras} {"Q. J. Roy. Astron. Soc."}
\newcommand{\skytel} {"Sky Telescope"}
\newcommand{\solphys} {"Sol. Phys."}
\newcommand{\sovast} {"Sov. Astron."}
\newcommand{\ssr} {"Space Sci. Rev."}
\newcommand{\zap} {"Zeitschrift fuer Astrophysik"}
\newcommand{\nat} {"Nature"}
\newcommand{\iaucirc} {"IAU Cirulars"}
\newcommand{\gca} {"Geochim. Cosmochim. Ac."}
\newcommand{\grl} {"Geophys. Res. Lett."}
\newcommand{\jcp} {"J. Chem. Phys."}
\newcommand{\jgr} {"J. Geophys. Res."}
\newcommand{\jqsrt} {"J. Quant. Spectrosc. RA"}
\newcommand{\nphysa} {"Nucl. Phys. A"}
\newcommand{\physrep} {"Phys. Rep."}
\newcommand{\physscr} {"Phys. Scrip."}
\newcommand{\planss} {"Planet. Space Sci."}
\newcommand{\baas} {"Bull. Aust. Acoust. Soc"}
\newcommand{\aplett} {"Astrophys. Lett."}
\newcommand{\procspie} {"Proc. SPIE"}
\newcommand{\cjaa} {"Chinese J. Astron. Ast."}
\newcommand{\fcp} {"Fundam. Cosm. Phys."}
\newcommand{\memsai} {"Mem. Soc. Astron. Ital."}
\newcommand{\bain} {"Bull. Astron. Inst. Neth., Suppl. Ser."}
\shorttitle{Ions in 67P/C-G}    

\shortauthors{Ahmed and Soni}  

\title [mode = title]{Modeling the Plasma Composition of 67P/C-G at different Heliocentric Distances}  



%

\author[1]{Sana Ahmed}[orcid=0000-0002-1258-6032]

\cormark[1]


\ead{ahmed.sana92@gmail.com}



\affiliation[1]{organization={Planetary Sciences Division, Physical Research Laboratory},
            addressline={University Area}, 
            city={Ahmedabad},
            postcode={380009}, 
            state={Gujarat},
            country={India}}

\author[1]{Vikas Soni}[orcid=0000-0001-9273-9694]






\cortext[1]{Corresponding author}



\begin{abstract}
\noindent The \textit{Rosetta} spacecraft accompanied the comet 67P/C-G for nearly 2 years, collecting valuable data on the neutral and ion composition of the coma. The Rosetta Plasma Consortium (RPC) provided continuous measurements of the in situ plasma density while ROSINA-COPS monitored the neutral composition. In this work, we aim to estimate the composition of the cometary ionosphere at different heliocentric distances of the comet. \cite{Lauter2020} derived the temporal evolution of the volatile sublimation rates for 50 separated time intervals on the orbit 67P/C-G using the COPS and DFMS data. We use these sublimation rates as inputs in a multifluid chemical-hydrodynamical model for 36 of the time intervals for heliocentric distances $< 3$ au. We compare the total ion densities obtained from our models with the local plasma density measured by the RPC instruments. We find that at the location of the spacecraft, our modeled ion densities match with the in situ measured plasma density within factors of $1-3$ for many of the time intervals. We obtain the cometocentric distance variation of the ions \ch{H2O+} and \ch{H3O+} and the ion groups created respectively by the ionization and protonation of neutral species. We see that \ch{H3O+} is dominant at the spacecraft location for nearly all the time intervals while ions created due to protonation are dominant at low cometocentric distances for the intervals near perihelion. We also discuss our ion densities in the context of their detection by DFMS. 

\end{abstract}





\maketitle
\section{Introduction} \label{sec:intro}
\noindent Comet 67P/Churyumov-Gerasimenko (henceforth 67P/C-G) is a Jupiter family comet that is currently characterized by an orbital period of 6.44 yr, and perihelion and aphelion distances of 1.24 au and 5.68 au, respectively. It was the target of the European Space Agency's \textit{Rosetta} mission \citep{Schulz2009} that orbited in its vicinity for nearly 26 months, starting from the first encounter on 6 August 2014 at a heliocentric distance of 3.6 au. The spacecraft subsequently escorted 67P/C-G through the 2-year period that included the comet's arrival at perihelion on 13 August 2015, and then post-perihelion up to the end of the mission on 30 September 2016. \textit{Rosetta} typically remained at cometocentric distances ranging from tens to hundreds of kilometers during the mission period.

The sublimation of volatile ices from the comet nucleus forms the coma. The neutral species in the coma are partially ionized due to photoionization, electron impact ionization, and charge exchange with the solar wind \citep{Cravens1987}, thereby forming the cometary ionosphere. Changes in the comet's heliocentric distance lead to alterations in the cometary outgassing rate and activity patterns. The continuous monitoring of 67P/C-G by the \textit{Rosetta} suite of instruments has enabled studies on the evolution of the coma for varying heliocentric distances and spatial locations, including seasonal variability. Among the instruments carried by \textit{Rosetta} was the Rosetta Orbiter Spectrometer for Ion and Neutral Analysis (ROSINA; \citealp{Balsiger2007}) that included the  Double Focusing Mass Spectrometer (DFMS), the Reflection-type Time-Of-Flight (RTOF) mass spectrometer, and the COmet Pressure Sensor (COPS). While ROSINA-DFMS was able to measure mass spectra in the range of 13 - 140 uq$^{-1}$ in low and high resolution modes, ROSINA-COPS provided us with time-series measurements of the total neutral density. \textit{Rosetta} was also equipped with the instruments of the Rosetta Plasma Consortium (RPC; \citealp{Carr2007}) for monitoring the cometary plasma parameters. RPC included the Mutual Impedance Probe (MIP; \citealp{Trotignon2007}) and the LAngmuir Probe (LAP; \citealp{Eriksson2007}) that were used to derive the in situ plasma density. 

Analyses of the ROSINA-DFMS spectra led to the detection of a large number of different ion species in 67P/C-G. The high resolution mode of DFMS ($m/\Delta m>3000$ at 1\% peak height for 28 uq$^{-1}$; \citealt{Balsiger2007}) could distinguish between ions that had very small difference in mass-per-charge ratios, e.g. \ch{H2O+} and \ch{NH4+}. \cite{Beth2016} report on the first in situ detection of \ch{NH4+} ions when 67P/C-G was near perihelion. An ionospheric model given by \cite{Vigren2013} that solves the continuity equation is used by \cite{Beth2016} to understand the chemistry of \ch{NH4+} ions for coma conditions encountered in July-August 2015. \cite{Beth2020} report on the in situ detections of cometary ions over the range $13-39$ uq$^{-1}$ from data acquired by DFMS in the high resolution mode and the first unambiguous detection of a dication (doubly charged ion) \ch{CO2^{++}} in a cometary ionosphere.

Previous studies have used a multi-instrument data-based ionospheric model to quantify the ionization sources in 67P/C-G \citep{Galand2016, Hajra2017, Heritier2017a, Heritier2018}. 
\cite{Galand2016} applied this model to the large heliocentric ($r_h > 3$ au) and low cometocentric distances ($r<20$ km) during the early mission period (October 2014) and found that solar EUV photoionization and electron impact ionization are the main ionization sources in the coma. \cite{Galand2016} also found variability in the ionization sources with spatial location; while photoionization dominated in the northern hemisphere (summer), the contribution of electron impact ionization had to be taken into account in order to explain the ion density over the southern hemisphere (winter). 
\cite{Heritier2018} used this model for post-perihelion cases ($r_h>2.6$ au; March-August 2016), and also found seasonal variability as \cite{Galand2016}. 
\cite{Heritier2018} also found that near perihelion, where the neutral density is the highest, photoionization is the dominant ionizing source, while electron impact ionization is dominant at large heliocentric distances. \cite{Hajra2017} applied the ionospheric model to study the plasma density during a cometary outburst in February 2016, while \cite{Heritier2017a} used it to study the near-surface cometary ionosphere during the final descent of \textit{Rosetta} at the end of the mission (September 2016).  \cite{SimonWedlund2019} found that the ionization due to solar wind charge exchange, though normally less than photoionization by factors of $5-100$, may sometimes become higher than the photoionization frequency during transient events such as interplanetary coronal mass ejections.


Although the total ion density can be known from the LAP and MIP measurements \citep{Edberg2015, Eriksson2017, Henri2017, Johansson2021}, it was difficult to directly measure the number density of each ionic species. The narrow field of view of DFMS could only sample a small part of the full ion distribution, and pickup by the solar wind could further deflect the ions from the DFMS field of view \citep{Fuselier2015}. 
In this work, we aim to estimate the composition of the ionosphere of 67P/C-G at different locations in its orbit around the Sun. For this purpose, we have obtained numerical solutions to a multifluid chemical-hydrodynamical model by varying the input conditions (volatile sublimation rates) at different orbital locations. Our model outputs give us the variation of the densities of assorted ions with cometocentric distance and we compare the total ion densities obtained from our models with the in situ measurements of LAP and MIP. Section \ref{sec:model} describes the modeling approach while Section \ref{sec:obs} contains a description of the instrument datasets and the input conditions which we have made use of in this study. The results and discussions are presented in Sections \ref{sec:res} and \ref{sec:dis}, respectively and the concluding remarks are made in Section \ref{sec:concl}.

\section{The Coma Model} \label{sec:model}
\noindent We use a combined chemical-hydrodynamical model suitable for a coma that is assumed to be spherically symmetric and in a steady state. The model philosophy is given by the multifluid work of \cite{Rodgers2002}, with subsequent additions given by \cite{Weiler2007Thesis} and \cite{Ahmed2021}. The gas phase coma is approximated as a fluid; a multifluid approach is adopted, in which the coma is divided into three fluids, namely the ions, the neutrals, and the electrons. These three fluids have individual temperatures and a common bulk velocity.
In addition to the creation of ions, radicals, and electrons in the coma due to photodissociation/photoionization, the species in the coma undergo a host of gas-phase chemical reactions, which include ion-neutral and neutral-neutral collisional reactions, dissociative recombination reactions, and electron impact reactions.
The energy released due to the chemical reactions is distributed non-uniformly among all of the species, resulting in different temperatures of the ions, neutrals, and electrons, and hence our multifluid approach. The expansion velocity for all species is assumed to be the same. Since the ions and electrons are coupled due to Coulomb interactions, resulting in charge neutrality, the assumption of a common velocity (plasma velocity) for the ion and electron fluids in a spherically symmetric coma is reasonable. In some regions of the coma, the electron temperature can attain very high values ($\sim10^4$ K), causing the plasma velocity to become subsonic. A smooth transition through the sonic point requires a separate numerical treatment. To avoid this, we assume that the plasma velocity is the same as the velocity of the neutral species so that all of the fluids move with a single bulk velocity \citep{Rodgers2002, Weiler2007Thesis, Holscher2015Thesis, Ahmed2021}.

Apart from heat exchange between fluids due to chemical reactions, there is also the exchange of energy by means of elastic and inelastic scattering (for example, ion-electron Coulomb interactions, electron-neutral ro-vibrational excitations of neutral species, and so on).  A full model description of all the components that need to be calculated to arrive at numerical solutions is given in \cite{Ahmed2021}. The modeling approach is based on obtaining numerical solutions to the equations for the conservation of number density, mass, momentum, and energy. The steady-state and spherically symmetric continuity equation for a species $i$ having number density $n_i$, bulk velocity $v$, production rate $P_i$, and loss rate $L_i$ at a cometocentric distance $r$, is given by:

\begin{equation}
    \frac 1 {r^2} \frac d {dr} (r^2n_iv) = P_i - L_i.
    \label{eq:N}
\end{equation}
The above equation is coupled with the differential equations for the conservation of mass, momentum, and energy, which are given by Equations \ref{eq:M}, \ref{eq:F} and \ref{eq:Q}, respectively.

\begin{equation}
\frac 1 {r^2} \frac d {dr} (r^2\rho v) = M,
\label{eq:M}
\end{equation}

\begin{equation}
\frac 1 {r^2} \frac d {dr} (r^2\rho v^2) +\frac d {dr} (n k_B T) = F,
\label{eq:F}
\end{equation}

\begin{equation}
\frac 1 {r^2} \frac d {dr} \left[ r^2\rho v\left(\frac {v^2} 2 + \frac \gamma {\gamma -1} \frac {k_B T} \mu \right )  \right]= E.
\label{eq:Q}
\end{equation}
In the above equations, $v$ is the common velocity for all fluids
and $k_B$ is the Boltzmann constant. $\rho$, $T$, $\gamma$, and $\mu$ are respectively the mass density, temperature, adiabatic 
exponent, and average molecular mass corresponding to each fluid (namely, neutral, ion, and electron). The source terms $M$, $F$ and $E$ represent the net rate 
per unit volume for the generation of mass, momentum, and energy, respectively for each fluid. We also have $\sum_{n,i,e} M = \sum_{n,i,e} F = 0$ (subscripts $n$, $i$ and $e$ respectively denote the neutral, ion and electron fluids). We define $G$ as the net rate of generation of thermal energy per unit volume for each fluid; we have $G_k = E_k-F_kv+ \frac 1 2 M_k v^2$, where $k \in \{n,i,e\}$. The components $G_k$ for each fluid are calculated as described in \cite{Ahmed2021}. The set of Equations $\ref{eq:N}-\ref{eq:Q}$ can be written as first-order differential equations which we numerically integrate for different gas production rates $Q$.

\subsection{Calculation of Ionization Rates}
\label{subsec:phot}
\noindent Photoionization due to the solar UV flux is a significant driver of the cometary ionospheric population. The UV radiation field does not remain constant throughout the coma; absorption of the UV photons by the cometary neutrals and scattering by nanograins in the coma leads to the attenuation of the UV flux, thus altering the rates of photochemical reactions at different cometocentric distances. The Beer-Lambert law can be used to derive the flux $\phi(\lambda, r)$ at any cometocentric distance $r$ and wavelength $\lambda$:

\begin{equation}
    \phi(\lambda, r)=\phi(\lambda,r_{\infty}) e^{-\tau(\lambda, r)}.
    \label{eq:photo}
\end{equation}
$\phi(\lambda,r_{\infty})$ is the spectral flux reaching the top of the coma, and the quantity in the exponential is the optical depth, such that:

\begin{equation}
    \tau (\lambda, r) = \sum_i \sigma_{i,\text{tot}}(\lambda) \int_r ^\infty  n_i(r^{\prime}) dr^{\prime}.
    \label{eq:tau}
\end{equation}
In the above expression, $n_i(r^{\prime})$ is the number density of a neutral cometary species $i$ that is present along the trajectory of the photon at a radial distance $r^{\prime}$, and $\sigma_{i,\text{tot}}(\lambda)$ is its total photo-absorption cross section at wavelength $\lambda$. This expression is valid for the one-dimensional case, i.e., when the spacecraft is along the sun-comet axis. Since that is not usually the case, the radial distance is written as $r=\left(x^2+y^2+z^2\right)^{1/2} $, where $x$ is along the sun-comet vector and $(y,z)$ depends on the position of the spacecraft. Equation \ref{eq:tau} is modified as follows:

\begin{equation}
\tau (\lambda, r) = \sum_i \sigma_{i,\text{tot}}(\lambda) \int_{x_0}^{\infty} n_i\left(\sqrt{x^2+y_0^2+z_0^2}\right) dx,
\end{equation}
where $(x_0, y_0, z_0)$ is the position vector of the spacecraft. The photochemical reaction rate $k(\lambda, r)$ of the species $i$ in the wavelength bin $\lambda$ and $\lambda + \Delta \lambda$ is:

\begin{equation}
    k(\lambda,r)=\sigma_{i,\text{br}}(\Delta\lambda) \Phi(\Delta\lambda,r).
\end{equation}
where  $\sigma_{i,\text{br}}(\Delta\lambda)$ is the wavelength-averaged partial photo-absorption cross section in the wavelength bin and $\Phi(\Delta\lambda,r)$ is the spectral photon flux obtained from Equation \ref{eq:photo} and integrated over $\Delta \lambda$. We sum over all the wavelength bins to get the total rate  coefficient (s$^{-1}$) for the photochemical reaction at a cometocentric distance $r$.

We have used the wavelength-dependent photo-absorption cross sections contained in the Photo Ionization/ Dissociation Rates (PHIDRATES) database, available at \href{https://phidrates.space.swri.edu/}{phidrates.space.swri.edu}. This database is compiled from cross sections given by \cite{Huebner1979}, \cite{Huebner1992} and \cite{Huebner2015}. There is no instrument on board \textit{Rosetta} for directly measuring the solar UV flux. \cite{Johansson2017} extracted the photoelectron saturation current from the RPC Langmuir probe measurements, which can be used as an index of the solar far and extreme ultraviolet at the location of \textit{Rosetta}. During the early and late mission phase of \textit{Rosetta}, this data set is well correlated with the UV observations by TIMED/SEE and MAVEN/EUVM, though there is a decrease in the expected photoelectron current near perihelion. The effect of this decrease on our model results is discussed further in Section \ref{sec:res}. In the present case, for the solar spectral UV photon flux data, we rely on the wavelength-dependent solar irradiance data set that is available with the Laboratory for Atmospheric and Space Physics Interactive Solar Irradiance Data Center (\href{https://lasp.colorado.edu/lisird/}{lasp.colorado.edu/lisird}). The spectral fluxes that we have used are derived from the FISM2 ($0.1-190$ nm) and NRLSSI2 ($>190$ nm) models. FISM2 \citep{Chamberlin2020} is an empirical model that is based on data from SORCE/XPS \citep{Woods2005}, SORCE/SOLSTICE \citep{Rottman2005}, and SDO/EVE \citep{Woods2012} while NRLSSI2 makes use of the solar irradiance measurements obtained by SORCE \citep{Harder2005, Kopp2005, Rottman2005}. The spectral fluxes given by FISM2 and NRLSSI2 are available at 1 au, and in order to scale them to some other heliocentric distance $r_h$, we have used the multiplicative factor $r_h^{-2}$, where $r_h$ is in units of au.

The ionization and dissociative ionization of neutral species also occur due to collisions with energetic electrons present in the coma. To calculate the ionization frequencies, we assume a Maxwellian distribution of the electrons. At an electron temperature $T_e$, this distribution is written as:

\begin{equation}
    f(v)=[m_e/2\pi k_B T_e]^{3/2} \exp{\{-m_e v^2/2k_BT_e\}},
\end{equation}
where $m_e$ is the electron mass and $k_B$ is the Boltzmann constant. The ionization frequency $R_{ik}$ (s$^{-1}$) for a species $i$ attaining an ionized state $k$ can be calculated as:

\begin{equation}
    R_{ik} = n_e \int_{v_{ik}}^{\infty} v\sigma_{ik}(v)f(v)4\pi v^2 dv, 
    \label{eq:eii}
\end{equation}
where $n_e$ is the electron number density, $v_{ik}$ is the velocity corresponding to the ionization potential $I_{ik}$ and $\sigma_{ik}$ is the cross section. The compiled cross sectional data for the electron impact ionization of various neutral species is available in the literature. These include cross sections for the ionization of the main cometary volatiles namely \ch{H2O} \citep{Itikawa2005}, \ch{CO2} \citep{Itikawa2002} and \ch{CO} \citep{Itikawa2015}, as well as for other cometary neutrals namely \ch{N2} \citep{Itikawa2006}, \ch{O2} \citep{Itikawa2009}, \ch{H2} \citep{Yoon2008}, and \ch{CH4} \citep{Song2015}. We use these cross sectional data to calculate the electron impact ionization rates from  Equation \ref{eq:eii}. Modeling of the instrument response of \textit{Rosetta}'s RPC-MIP shows the presence of two electron populations that follow a double Maxwellian electron velocity distribution function \citep{Gilet2017, Wattieaux2019}. These two populations are hot electrons of energy $5-15$ eV and cold electrons of energy $0.1 - 1$ eV. The cold electron population does not possess sufficient energy to ionize neutral molecules (ionization threshold energy of cometary neutrals is $> 12$ eV). Thus, calculating the ionization frequency using a single Maxwellian distribution for the electrons is an acceptable approximation. 

\section{Dataset} \label{sec:obs}
\subsection{The Rosetta Plasma Consortium} \label{sub:rpc_all}
\noindent The five sensors on the Rosetta Plasma Consortium (RPC) are the Ion and Electron Sensor (IES; \citealt{Burch2007}), the Ion Composition Analyzer (ICA; \citealt{Nilsson2007}), the Langmuir Probe (LAP; \citealt{Eriksson2007}), the Mutual Impedance Probe (MIP; \citealt{Trotignon2007}) and the Magnetometer (MAG; \citealt{Glassmeier2007}). IES, ICA, LAP and MIP provide measurements of the plasma number density and each instrument probes different plasma populations (e.g., cold/warm electrons, energetic electrons, cometary ions) associated with different energy ranges. RPC-IES is an electrostatic analyzer designed to measure the ion and electron flux over the energy range from 4.32 eV q$^{-1}$ to 17.67 keV q$^{-1}$. RPC-ICA is an ion spectrometer that scans the energy and angular space to detect ions of different masses. Its operational energy range is a few eV q$^{-1}$ up to 40 keV q$^{-1}$ and it can distinguish between \ch{H+}, \ch{He+}, \ch{He^2+} and heavy cometary ions of mass corresponding to water group ions and above. 
	
The RPC-ICA number density measurements are best in low density plasmas such as solar wind or cometary ions at large cometocentric distances. A limitation of this sensor is that the negative spacecraft potential complicates the observations of ions with low energies. Generally only a fraction of the low energy ions are seen by ICA, and the cometary ion density estimates are lower than those from LAP and MIP. There is a net anti-sunward flow of cometary ions \citep{Nilsson2015, Nilsson2017, Bercic2018}, and ICA mainly observes the accelerated ions while the more locally produced plasma is below the threshold energy level of ICA. The negative spacecraft potential affects the detection capability of RPC-IES as well, since low energy electrons cannot reach the sensor and the electron number density is underestimated. In this work, we use the plasma densities measured from RPC-LAP and RPC-MIP; these are described in the following section.

\subsection{Plasma Density from LAP and MIP} \label{sub:rpc}
\noindent RPC-LAP consists of two spherical Langmuir probes mounted at the tips of two solid booms that protrude non-symmetrically from the spacecraft \citep{Eriksson2007}. A standard mode of operation of the Langmuir probe is the bias voltage sweep. During a sweep, the probe bias voltage with respect to the spacecraft is stepped through a series of values, and the current collected by the probe is measured. The current flowing in the probe is proportional to the plasma density and also depends on other factors such as the plasma energy distribution, the applied bias voltage, and the spacecraft potential. The total probe current can be separated into three parts, namely ion, electron, and secondary electron emission currents \citep{Eriksson2017, Johansson2017}. Electrons dominate the collected current when the probe is at a positive potential with respect to the plasma and the electron number density can be determined from this part of the probe current. When the potential is negative, positively charged ions are collected. In a tenuous and sunlit plasma, the photoelectrons emitted from the probe dominate the collected current, which can be used to estimate the integrated solar UV flux \citep{Johansson2017}. 

RPC-MIP is an active electric sensor that measures the transfer impedance between a transmitting monopole or dipole and a receiving dipole \citep{Trotignon2007}. The instrument offers different time resolutions and operates within different frequency bands in the $7-3500$ kHz range. There are two modes of operation, namely the passive mode in which the transmitters are off, and the active mode in which the transmitters stimulate the surrounding plasma. When operating in the active mode, the electrodes on the RPC-MIP can be used independently or cojointly as monopole or dipole transmitters, respectively. This is known as the SDL (Short Debye Length) mode, while in the LDL (Long Debye Length) mode, the LAP2 probe of RPC-LAP serves as the monopole transmitter. The LDL mode, while preventing RPC-LAP from being fully operational, enables the measurement of plasma of lower density than that measured in the SDL mode since plasma at larger spatial distances from the receivers is triggered. Since charge neutrality is maintained, the ion density in the plasma can be estimated from the electron density. RPC-MIP measures the electron density by identifying the plasma resonance frequency $f_p$ in the mutual impedance spectrum, i.e., the response of the plasma to a weak transmitted electric signal. $f_p$ is only dependent on the number density and thus, MIP can provide accurate density estimates within its operational range. Due to instrumental limitations, RPC-MIP can measure the electron density in the range of $10-10^5$ cm$^{-3}$ for energy (eV) $<0.05$ $n_e$ in the SDL mode and $1-350$ cm$^{-3}$ for energy (eV) $<0.15$ $n_e$ in the LDL mode.  

If the electron density is too low in the SDL mode of RPC-MIP, the density can be retrieved from the LDL mode, though the LDL mode has its limitations as well. If the Debye length is too high (for low electron density), the instrument becomes blind to the plasma. On the other hand, if the plasma frequency is higher than the detectable limit (for high electron density), saturation is reached in the measured density \citep{Heritier2018}. The dataset obtained by the cross-calibration of the measurements of RPC-LAP and RPC-MIP overcomes these operational limitations and is available in ESA's Planetary Science Archive \citep{Besse2018}. The recommended dataset for long-term studies is the low time resolution LAP/MIP cross-calibrated data product known as NED, available as ned.tab files in the LAP derived data folder in PSA. This dataset has a time resolution varying between 32 s and a few minutes, a high dynamic range, and the broadest coverage over the mission period, and its homogeneous nature makes it suitable for statistical studies. The accuracy of an individual data point is less than the MIP density, but the wide dynamic range ensures lesser systematic bias towards high or low densities.  

\subsection{Neutral Density from ROSINA} \label{sub:rosina}
\noindent The ROSINA-COPS pressure sensor consists of two gauges, namely the nude gauge that measures the total density of the cometary gas at the spacecraft location and the ram gauge that measures the ram pressure which is equivalent to the gas flux. The COPS operating principle is that the gas is ionized by electron impact and the resulting ion current is measured. The instrument sensitivity to the different gas species is not the same due to differences in the electron impact ionization cross sections and correction factors are required for the individual species. The relative abundances of the species in the gas, available from the DFMS and RTOF measurements, are required to calculate the weighted average of the ionization cross section and the absolute abundances of the species \citep{Gasc2017}. The level 4 ROSINA density data in PSA is the corrected data that represents the actual neutral density of the coma. 

\cite{Lauter2020} considered the period from 1 August 2014 (377 days before perihelion) to 5 September 2016 (390 days after perihelion) and divided this period into 50 time intervals of length varying between 7 to 29 days. The intervals $I_{1-50}$ were chosen in a way that the sub-spacecraft position samples sublimation from nearly all the surface elements of 67P/C-G, but there is limited variation in the heliocentric distance and subsolar latitude within the interval. For each of these intervals, \cite{Lauter2020} used the combined DFMS and COPS measurements to derive the temporal evolution of the production rates of 14 volatile species belonging to the set $S$ where 

\begin{equation}
    S = \left\{ \ch{H2O}, \ch{CO2}, \ch{CO}, \ch{H2S}, \ch{O2}, \ch{C2H6}, \ch{CH3OH}, \ch{H2CO}, \ch{CH4}, \ch{NH3}, \ch{HCN}, \ch{C2H5OH}, \ch{OCS}, \ch{CS2} \right\}.
    \label{set:S}
\end{equation}
These production rates compare well with other production rate results calculated based on the \textit{Rosetta} instruments VIRTIS \citep{Fougere2016}, MIRO \citep{Marshall2017, Biver2019}, DFMS \citep{Combi2020} and COPS \citep{Hansen2016}. 
\begin{figure}[t!]
    \centering
    \includegraphics[width = \textwidth]{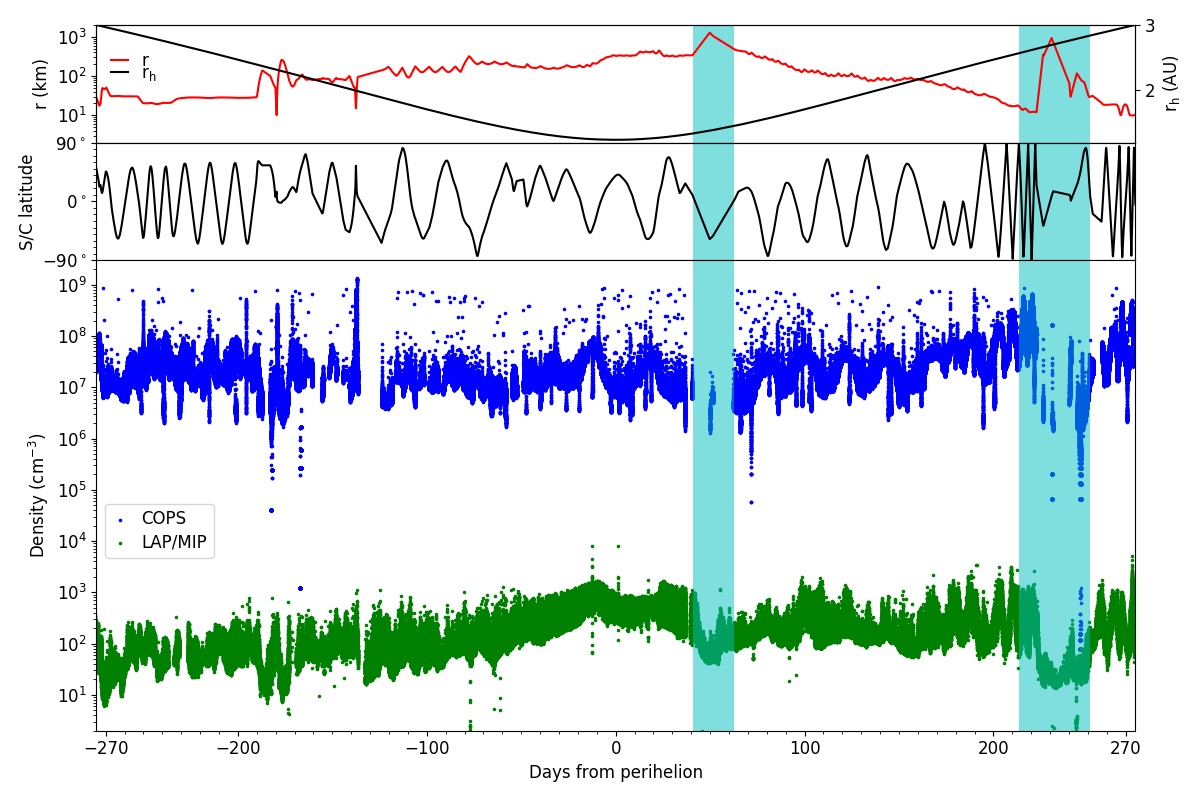}
    \caption{Top panel: time series of the cometocentric distance of \textit{Rosetta} (red line, left y-axis) and the heliocentric distance of 67P/C-G (black line, right y-axis). Middle panel: time series of the sub-spacecraft latitude. Bottom panel: time series of the neutral density measured by COPS (blue pixels) and the cross-calibrated electron density from LAP/MIP (green pixels). The shaded vertical regions indicate the time period around the dayside and tail excursions.
    }
    \label{fig:neutral-plasma}
\end{figure}

\subsection{Model Runs}
\noindent The bottom panel in Figure \ref{fig:neutral-plasma} shows the time series variation of the COPS neutral density and the LAP/MIP cross-calibrated electron density in the plasma. We use the convention of negative values for indicating days before perihelion. 
COPS data is unavailable or has large uncertainty during the "safe mode" of operations (1-10 April 2015) and the excursions to large cometocentric distances (dayside excursion in late-September to mid-October 2015 and tail excursion in late-March to early April 2016). The vertical shaded regions in Figure \ref{fig:neutral-plasma} indicate the time period around the excursions which have been excluded in the analysis by \cite{Lauter2020}; the time period of the safe mode was included in $I_{1-50}$. 

The comet surface can be treated as a mesh of triangular faces, each of which is treated as a source of volatile emission. The neutral gas density at the spacecraft location is the result of the superposition of gas expansion from these emission sources \citep{Kramer2017, Lauter2019}. Considering gas emission only from the surface (though subsurface sublimation may also occur), \cite{Lauter2020} retrieved the surface emissions using a mesh of 3996 triangular faces given by \cite{Preusker2017}. The time series of the retrieved production rates by \cite{Lauter2020} for all 14 volatiles in set $S$ is available as ancillary files to \href{https://arxiv.org/abs/2006.01750}{arxiv.org/abs/2006.01750}. We have used these production rates as inputs in our model runs. 
\section{Results} \label{sec:res}
\begin{figure}[htb!]
	\centering
	\includegraphics[width=\textwidth]{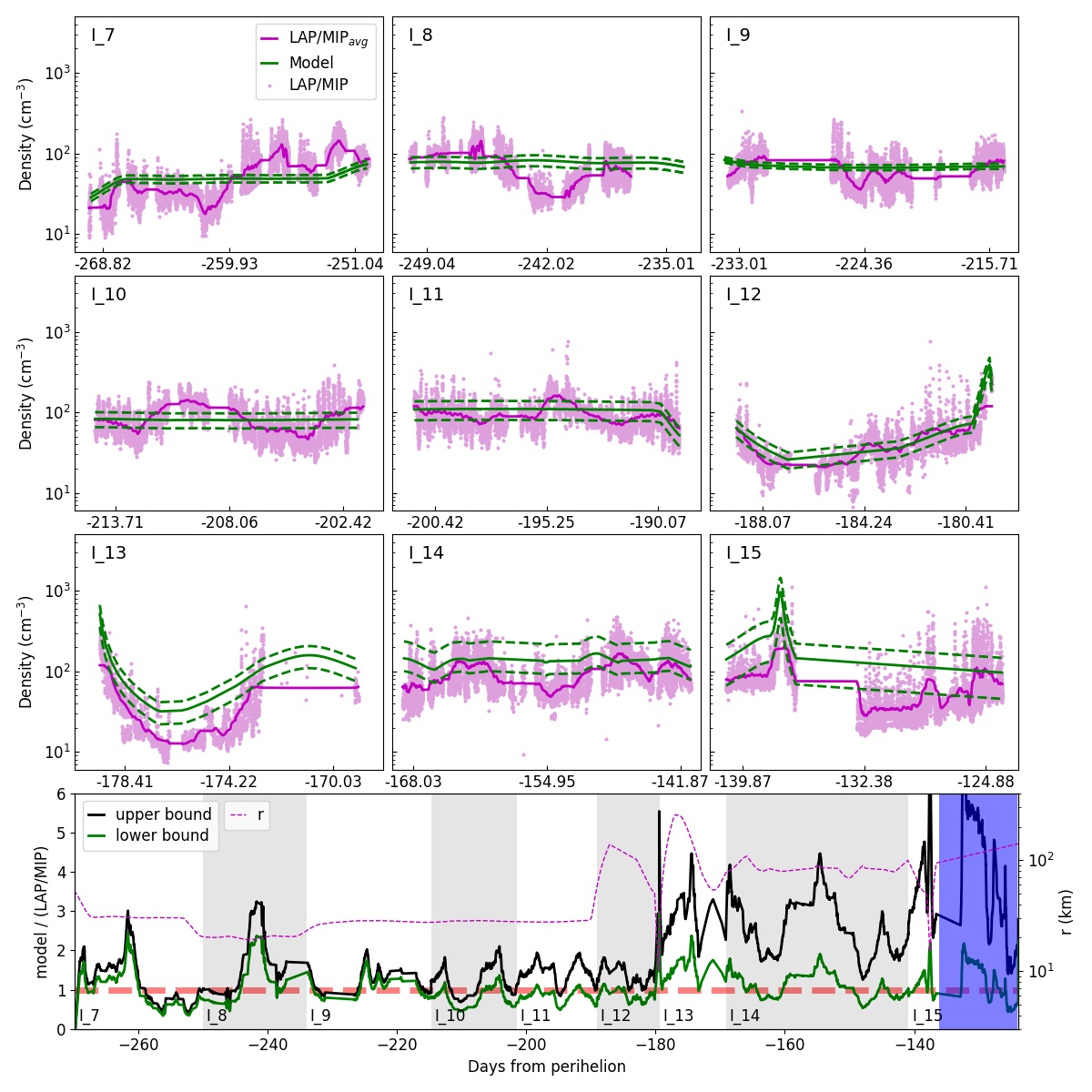}
	\caption{Subpanels in the top 3 rows: time series of the plasma density at the spacecraft location for the intervals $I_7$ to $I_{15}$. The pink pixels show the instantaneous LAP/MIP cross-calibrated density while the pink lines show the smoothed density obtained by taking a moving average. The green lines show the total ion density obtained from our models for the upper and lower bounds of the production rate (dashed lines) and the average of the upper and lower bounds (solid lines). Bottom panel: ratio between the modeled total ion density and the LAP/MIP plasma density for the upper (black line) and lower (green line) bounds of the production rate. The ratio is 1 on the dashed horizontal line. The intervals $I_7$ to $I_{15}$ are indicated by the white (odd-numbered intervals) and grey (even-numbered intervals) regions. The blue vertical shaded region indicates that the spacecraft was in the safe mode. The pink dashed line is the spacecraft distance from the nucleus (scale: right y-axis). }		
	\label{fig:ion1}
\end{figure}

\noindent We present our model results for the intervals in which the heliocentric distance $r_h < 3$ au (time intervals $I_7$ to $I_{42}$ as defined by \citealt{Lauter2020}). At large heliocentric distances, particularly when $r_h>3$ au, the outgassing rate is weak. Ions that are created due to the ionization of neutral species suffer only a limited number of collisions as the mean free path is larger at low densities, and the advection timescale is smaller than the chemical timescale \citep{Beth2016}. In each time interval, we have used our chemical-hydrodynamical model and obtained numerical solutions to Equation \ref{eq:N} to get the variation of the ion density with cometocentric distance. 

Due to limited surface coverage of the spacecraft, some of the surface elements do not have production rates assigned to them. To constrain the production rates from these elements, \cite{Lauter2020} have given upper and lower limits in their derived time series global production rates. The lower limit is set by assigning zero to the emission from the unknown surface element. The upper limit in the time interval $I_j$ is set by assigning the maximum production from that surface in the neighboring time intervals $I_{j-1}$ and $I_{j+1}$. Thus, in order to better constrain the modeled ion densities, we obtain two sets of model runs for each time interval and the input production rates are set as the upper and lower bounds respectively. We also obtain a third set of models where we take the average of the upper and lower bounds as inputs.

\subsection{Total Ion Density}
\begin{figure}[htb!]
	\centering
	\includegraphics[width=\textwidth]{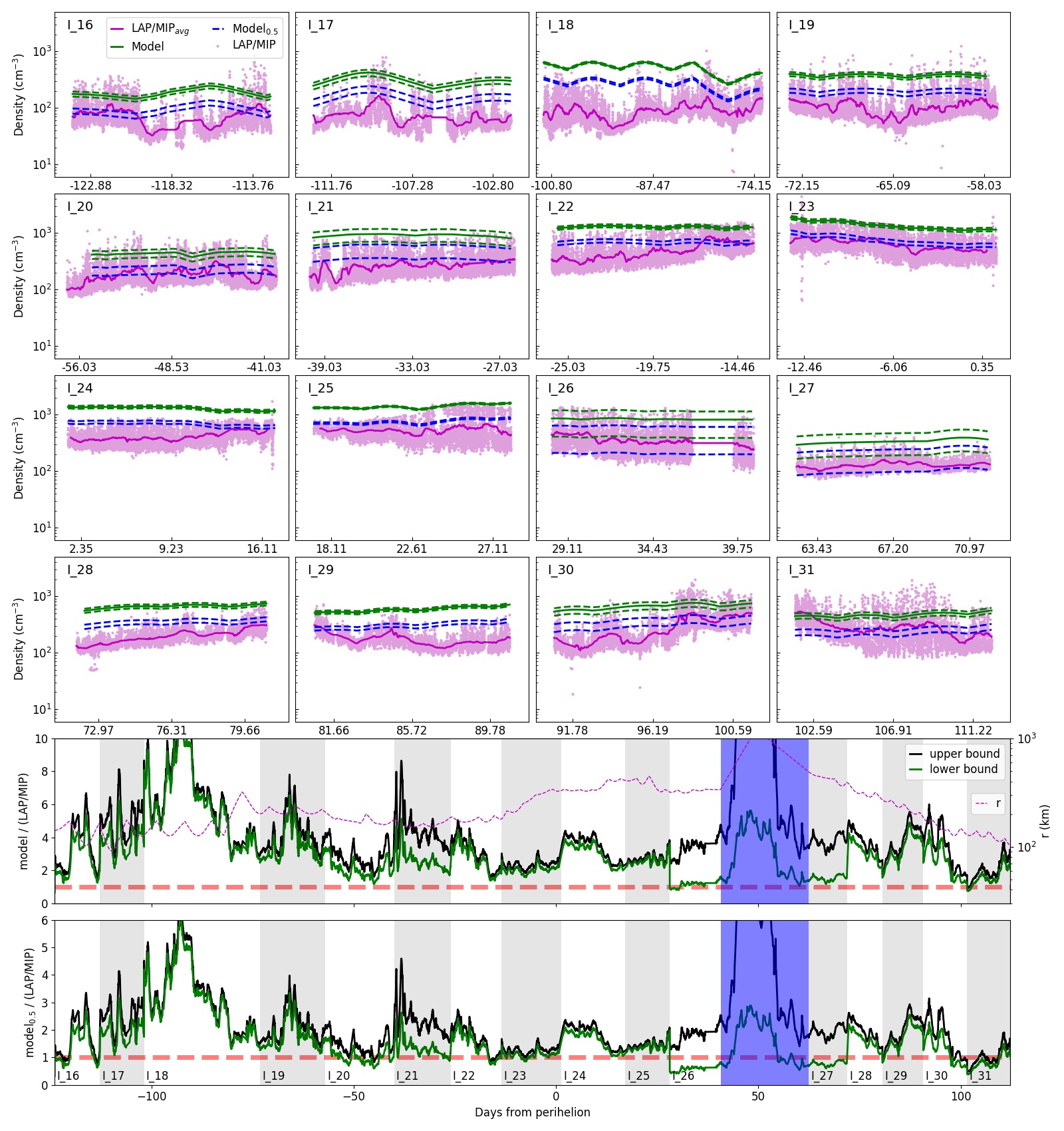}
	\caption{Same as Figure \ref{fig:ion1} but for the time intervals $I_{16}$ to $I_{31}$ and the following modifications: the blue dashed lines in the subpanels show the ion density for the upper and lower bounds of the production rate when the photoionization rate is reduced by 50\%. Of the two bottom panels, the upper one shows the density ratio described in the caption to Figure \ref{fig:ion1} while the lower one shows the density ratio at the reduced photoionization rate. The blue vertical shaded regions in the two lower panels indicate the dayside excursion.}		
	\label{fig:ion2}
\end{figure}

\begin{figure}[htb!]
	\centering
	\includegraphics[width=\textwidth]{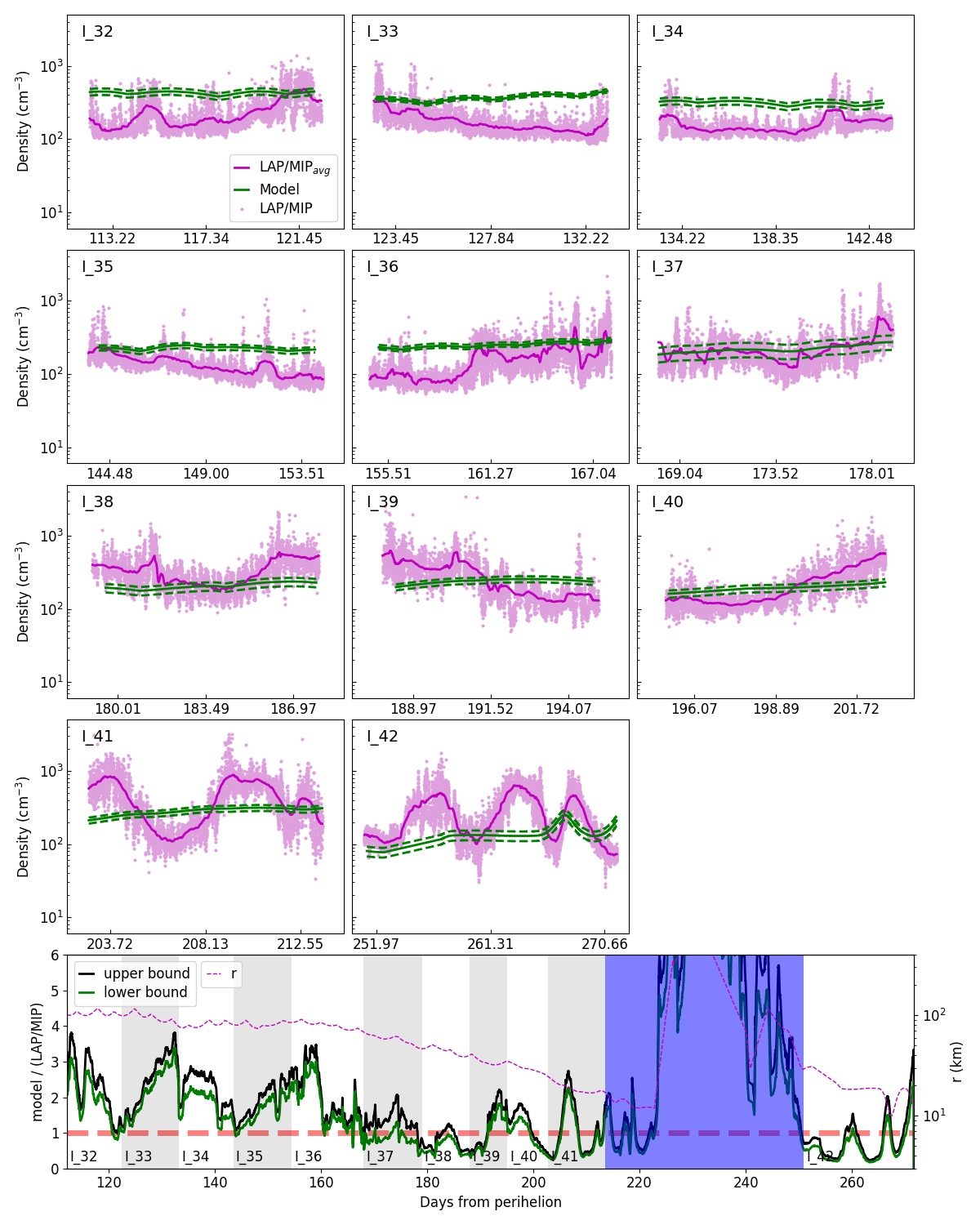}
	\caption{Same as Figure \ref{fig:ion1} but for the time intervals $I_{32}$ to $I_{42}$. The blue vertical shaded region in the bottom panel indicates the tail excursion. }		
	\label{fig:ion3}
\end{figure}

\noindent Figures \ref{fig:ion1}, \ref{fig:ion2}, and \ref{fig:ion3} show a comparison of our model obtained ion densities with the LAP/MIP cross-calibrated plasma density for different time intervals $I_j, j=(7,42)$. The LAP/MIP density data is smoothed by taking a moving average. In a particular time interval $I_j$, our model runs give the radial variation of the ion number density and we can find the number density at the cometocentric distance that corresponds to the spacecraft location. The model is not able to capture the high frequency density fluctuations in the plasma. The variations in our modeled ion density in the interval $I_j$ result due to changes in the spacecraft distance. We divide the time intervals into three groups in order to assess the trends in the modeled density.

\subsubsection{Group 1: Intervals $I_{7}$ to $I_{15}$}
\noindent The first group includes the intervals $I_{7}$ to $I_{15}$ (Figure \ref{fig:ion1}) and corresponds to the pre-perihelion period where $r_h>1.8$ au, up to the time when \textit{Rosetta} enters the safe mode. The total outgassing rate $Q$ varies from $\sim 10^{26}$ s$^{-1}$ to $< 10^{27}$ s$^{-1}$. In the first half of this group, there is limited change in the cometocentric distance and the spacecraft remains roughly at a height of $20-30$ km from the surface. In the latter part of this group, $r$ shows variation within a time interval, even becoming more than 100 km. This variation is reflected in the modulation of the ion density in $I_{13}$ and $I_{14}$, and a peak-like feature in the early part of $I_{15}$. Since the outgassing rate is low, most of the ions are created at smaller $r$ and a change in the cometocentric distance by 50 km or more is marked by a reduction in the total ion density. 

In order to estimate the extent to which our modeled densities deviate from the LAP/MIP observed plasma density, we have calculated the ratio between these two quantities, as shown in the bottom panel of Figure \ref{fig:ion1}. We see that for the lower bound of the production rate, this ratio remains close to 1. The fluctuations in the ratio are due to the fluctuations in the LAP/MIP density, which cannot be captured in our model as mentioned previously. For the upper bound of $Q$, the ratio mostly remains between $1-2$. In the intervals $I_{12}$ to $I_{15}$, the ratio for the lower bound of $Q$ varies between $1-2$, while the upper bound of $Q$ leads to an overestimation of the ion density by factors of $2-4$ (excluding the safe mode). 

The interaction of 67P/C-G with corotating interaction regions (CIRs) and coronal mass ejections lead to disruptions in the cometary plasma environment \citep{Edberg2016a, Edberg2016b, Hajra2017, Goetz2019}. Of the four CIRs that impacted 67P/C-G from October to December 2014, two of them, namely event 3 and event 4 defined by \cite{Edberg2016b} fall within the intervals $I_7$ and $I_9$, respectively. An increase in the observed density towards the end of $I_7$ corresponds to event 3 while the moderate increase in the density at the start of $I_9$ indicates the time of event 4. The response of the cometary plasma to individual events is variable, though these effects are not included in our model. 

\subsubsection{Group 2: Intervals $I_{16}$ to $I_{31}$}
\noindent The second group consists of the intervals $I_{16}$ to $I_{31}$ and the corresponding densities are shown in Figure \ref{fig:ion2}. This group includes the time intervals corresponding to the inbound leg from 1.8 au of the journey of 67P/C-G on its orbit around the Sun, the perihelion crossing at $r_h=1.24$ au, and the outbound journey post-perihelion up to 1.8 au. The spacecraft distance from the comet varies between $100-400$ km and the production rate ranges between $10^{27}$ s$^{-1} < Q < 10^{29}$ s$^{-1}$. In the intervals $I_{16} - I_{18}$, there is some modulation in the modeled ion density due to changes in $r$, while the density remains nearly constant in other intervals. The intervals near perihelion, characterized by high outgassing rates, exhibit ion densities exceeding 1000 cm$^{-3}$, even though $r>200$ km. The production rate calculated by \cite{Lauter2020} peaks in $I_{25}$ and we find that the total ion density is also maximum in this interval.

\cite{Johansson2017} used measurements from TIMED/SEE at Earth and MAVEN/EUVM at Mars for epochs when these two planets were at the same solar longitude as \textit{Rosetta} in order to estimate the flux propagated out to the comet location. They find that the photoemission current derived from TIMED/SEE and MAVEN/EUVM correlate well with the photoemission current obtained from the RPC-LAP measurements at 67P/C-G. They also find that around perihelion, the photoemission current from RPC-LAP reduces to nearly half of the expected value as estimated from MAVEN/EUVM due to attenuation of the UV flux. As described in Section \ref{subsec:phot}, our model includes the effect of the extinction of UV due to photo-absorption by the coma gas, though we do not find any appreciable change in the UV flux at the location of the spacecraft. \cite{Johansson2017} calculate the EUV absorption by water molecules to be only $0.8 \pm 0.1 \%$ near perihelion, at a cometocentric distance of 330 km. \cite{Heritier2018} also note that during the entire escort phase, there is no significant change in the plasma density due to photo-absorption. \cite{Johansson2017} propose scattering and absorption by cometary dust grains as a possible reason for the reduction in the photoemission current. Thus, we also ran models where we reduced the photoionization rate by half, and the resulting total ion densities are indicated by the blue dashed lines in the subpanels of Figure \ref{fig:ion2}. The total ion density at the reduced photoionization rate is closer to the LAP/MIP observed total plasma densities (Figure \ref{fig:ion2}, bottom panel). The ion ratio varies between $1-2$ for the lower bound of $Q$, and between $1-3$ for the upper bound. However, in $I_{18}$, the ratio reaches a high value, even at the reduced photoionization rate. \cite{Nemeth2020} finds that variations in the external solar wind pressure induce movements in the diamagnetic cavity boundary. For an inward moving boundary, the plasma density and suprathermal ion counts increase rapidly. Conversely, when the boundary moves outward, these quantities decrease, but more gradually. Thus, the decrease in the observed plasma density in $I_{18}$ (resulting in an increased model to LAP/MIP ion ratio) could be due to this effect. The observed plasma density modulations  due to solar wind pressure changes may also be present in the other intervals near perihelion ($I_{16} - I_{21}$). 

\subsubsection{Group 3: Intervals $I_{32}$ to $I_{42}$} 
\noindent The third group (Figure \ref{fig:ion3}) includes the intervals $I_{32}$ to $I_{42}$ where the heliocentric distance is $> 1.8$ au and $<3$ au. The spacecraft distance remains below 100 km, except during the tail excursion (not included in our model runs). The outgassing rate is $\sim 10^{27}$ s$^{-1}$ in the first half of the group, while it goes below $10^{26}$ s$^{-1}$ in the last few intervals. The decrease of \ch{H2O} post-perihelion is steeper than its increase pre-perihelion \citep{Lauter2020}. This is in contrast to volatiles such as \ch{CO} and \ch{CO2}, which show a slower decay at $r_h>2.4$ au.  Thus, there is a marked increase in the relative abundance of volatiles with respect to water; in fact, $Q_{\ch{CO2}}> Q_{\ch{H2O}}$ in the interval  $I_{42}$.

The ion ratios mostly vary between $2-3$ in the first half of this group. This corresponds to the period of December 2015 to late January 2016, for which \cite{Johansson2017} calculated the photoionization rates to be reduced by a factor of $\sim 0.6$. From Figure \ref{fig:ion2}, we see that a reduction in the photoionization rate by a factor of 0.5 leads to a reduction in the ion density by approximately the same amount. By applying a reduction of a factor of 0.6 to the intervals $I_{32}$ to $I_{36}$, the ion ratio would be within $1-2$. $I_{41}$ and $I_{42}$ show oscillation in their LAP/MIP measured density, which is correlated to the latitudinal variation of \textit{Rosetta}, and the ion ratio shows a similar oscillation. \cite{Nemeth2020b} defined a simple distance, latitude and longitude dependent first order cosine function to model the 3D spatial distribution of the cometary plasma. This model was able to reproduce quite well the plasma density distribution in the last month of \textit{Rosetta}'s operations. The features in $I_{41}$ and $I_{42}$ can probably be similarly explained. In $I_{42}$, the spacecraft is very close to the nucleus, the production rate is low, and our model underestimates the total ion density. 

\subsection{Density of Major Ions}

\begin{figure}[b!]
	\centering
	\includegraphics[width=\textwidth]{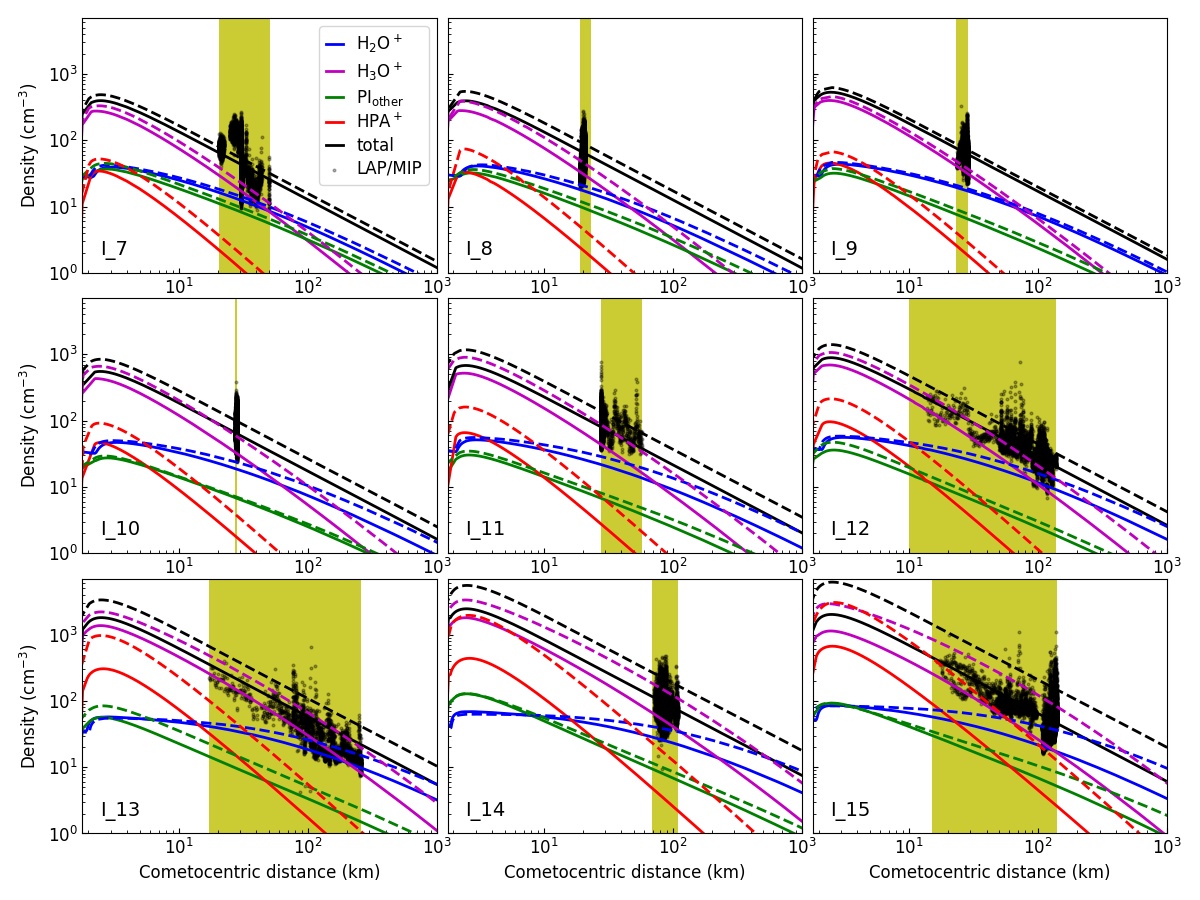}
	\caption{The subpanels show the variation of the total plasma density and the densities of \ch{H2O+}, \ch{H3O+}, \ch{HPA+}, and PI$_{\text {other}}$ ions with cometocentric distance for the time intervals $I_{7}$ to $I_{15}$. The solid lines and dashed lines denote the lower and upper bounds, respectively. The black pixels show the LAP/MIP density and the yellow vertical shaded regions in each subpanel denote the variation of the cometocentric distance of \textit{Rosetta} within the interval. }		
	\label{fig:n1}
\end{figure}

\begin{figure}[b!]
	\centering
	\includegraphics[width=\textwidth]{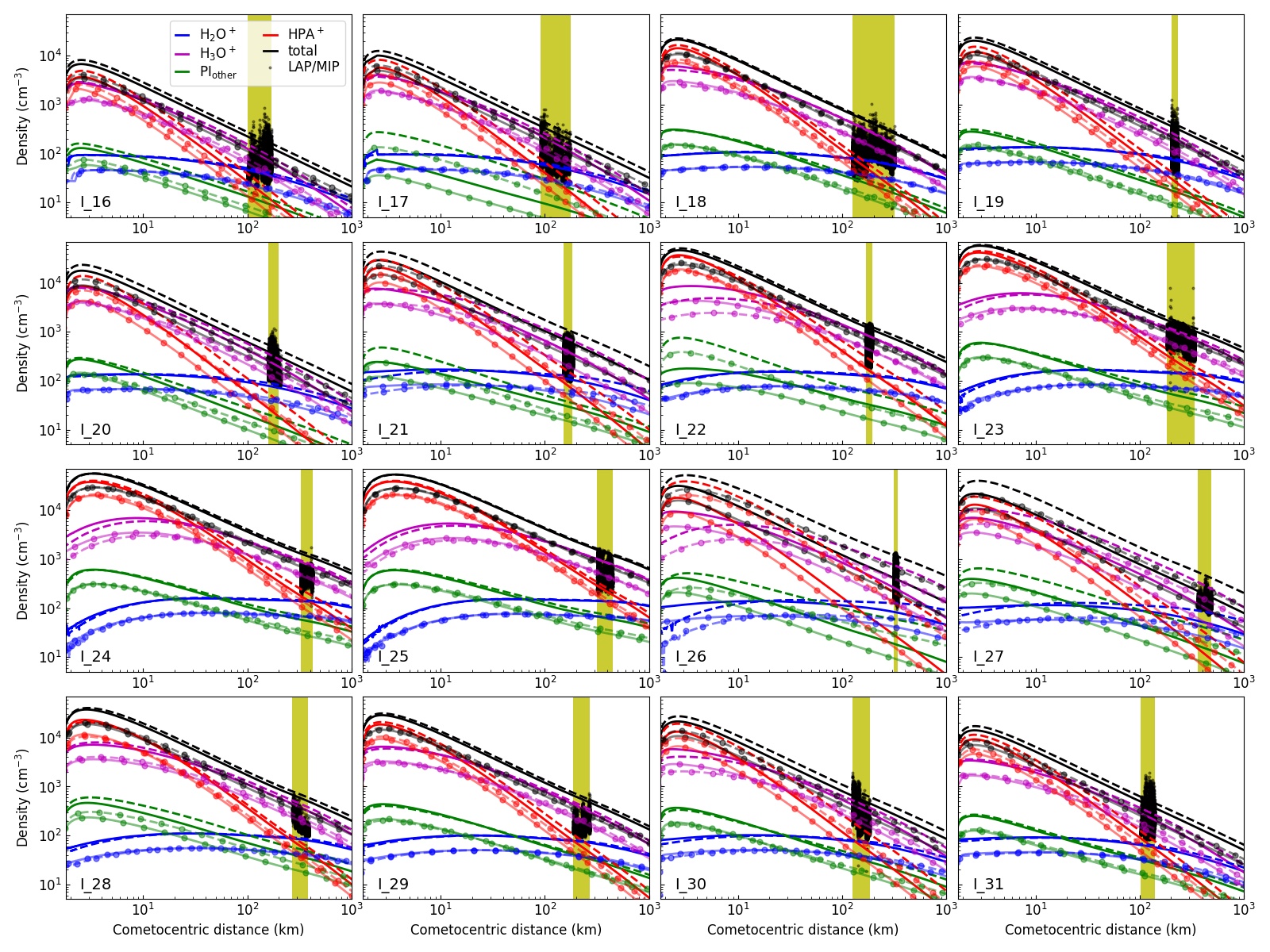}
	\caption{Same as Figure \ref{fig:n1} but for the time intervals $I_{16}$ to $I_{31}$. Additionally, the lines with circular markers denote the ion densities when the photoionization rate is reduced by 50\%. }		
	\label{fig:n2}
\end{figure}

\begin{figure}[b!]
	\centering
	\includegraphics[width=\textwidth]{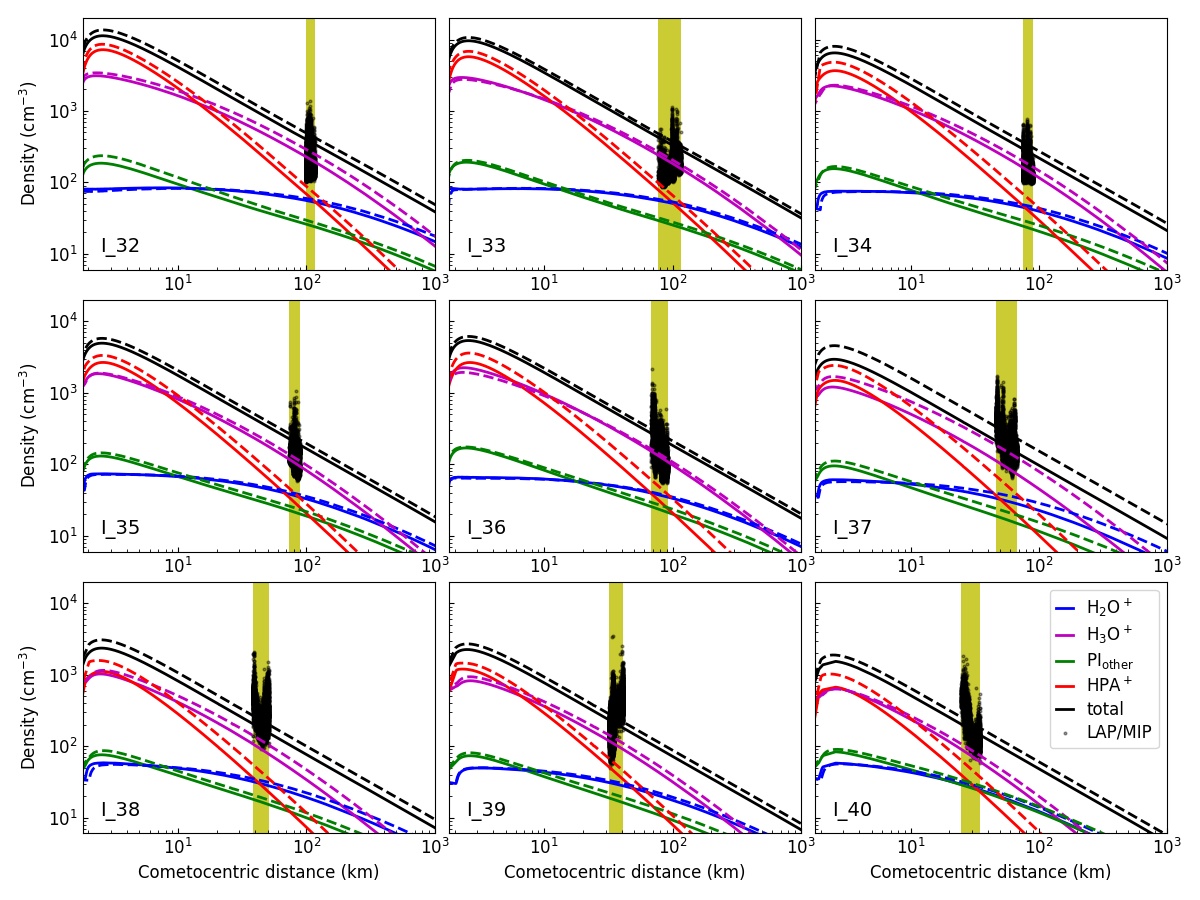}
	\caption{Same as Figure \ref{fig:n1} but for the time intervals $I_{32}$ to $I_{40}$.}		
	\label{fig:n3}
\end{figure}

\noindent In this section, we discuss the densities of some of the primary ions that form in the coma due to gas phase chemistry, as obtained from our model runs. The ions we consider are \ch{H2O+}, \ch{H3O+}, and the two ion groups labeled \ch{HPA+} and PI$_{\text {other}}$ in Figures $\ref{fig:n1} - \ref{fig:n3}$. \ch{H2O+} ions are created by the ionization of water molecules, which further react with \ch{H2O} to form \ch{H3O+} ions. It is well known that in the coma, a neutral species \ch{A} that has a proton affinity higher than water undergoes protonation by reacting with \ch{H3O+} ions to form \ch{A-H+} ions (see, for example, \citealt{Vigren2013, Heritier2017b}). From the parent neutrals in set $S$, we identify a subset that consists of these high proton affinity (HPA) neutrals, such that HPA = \{\ch{NH3}, \ch{CH3OH}, \ch{HCN}, \ch{H2CO}, \ch{H2S} and \ch{C2H5OH}\}. We define \ch{HPA+} as the sum of the ions (\ch{NH4+}, \ch{CH3OH2+}, and so on) formed by the protonation of the HPA neutrals. Since \ch{NH3} has the highest proton affinity, the terminal ion in the protonation chain is \ch{NH4+}, and \ch{NH3} will also accept \ch{H+} ions from other protonated species, such as \ch{CH3OH2+} and \ch{H3S+}. In general, a neutral species will undergo proton transfer reactions with \ch{H3O+} as well as other species that lie below it on the proton affinity ladder. Finally, PI$_{\text {other}}$ denotes the sum of the ionized parent species of set $S$ excluding \ch{H2O}, namely the sum of the ions \ch{CO2+}, \ch{CO+}, and so on. The cometocentric distance variation of \ch{H2O+}, \ch{H3O+}, \ch{HPA+} and PI$_{\text {other}}$ are shown in Figures $\ref{fig:n1}-\ref{fig:n3}$ for each of the intervals $I_j, j=(7,42)$.  The yellow-shaded regions show the extent of the change in the cometocentric distance of \textit{Rosetta} within each interval, indicating that the model estimated density of the ions lies within that region.

Following the same grouping of time intervals as described in the previous section, we see that in the first half of the group 1 intervals, \ch{H3O+} is the most abundant ion at the spacecraft location, followed by \ch{H2O+}. Since the coma density is low, \ch{HPA+} ions may form close to the nucleus surface, though their density decreases rapidly such that at the spacecraft location, the density $\lesssim 2$ cm$^{-3}$, which is less than what we estimate for \ch{H2O+} and PI$_{\text {other}}$ ions. As the comet moves closer to perihelion and there is an increase in the production rate, the density of \ch{HPA+} ions increases and may even become more than that of \ch{H2O+} for certain spacecraft locations. The density of PI$_{\text {other}}$ remains less than that of \ch{H2O+} at the spacecraft location since the sum of parent volatiles other than water makes up about $10-15\%$ of the total gas production. Out of these, the most dominant volatiles are \ch{CO2}, \ch{CO} and \ch{O2}, hence the main contributing ions are \ch{CO2+}, \ch{CO+} and \ch{O2+}. Apart from ionization, parent species also undergo dissociative ionization to produce ion fragments, for example, the dissociative ionization of \ch{CH4} creates \ch{CH3+}, \ch{CH2+} and \ch{CH+} ions, and \ch{NH3} produces \ch{NH2+} and \ch{NH+} ions. However, we do not consider these fragment ions in the present case, since their contribution to the total ion density at the spacecraft location is generally orders of magnitude lower than \ch{H2O+} \citep{Beth2020}. 

For the group 2 time intervals (Figure \ref{fig:n2}), the radial variation of the number density of \ch{H2O+} is relatively weaker than the group 1 intervals. The loss of \ch{H2O+} due to ion-neutral reactions is nearly equal to its rate of creation by ionization, causing it to remain in photochemical equilibrium. The increased outgassing rate drives more active chemistry and the \ch{HPA+} ions dominate over \ch{H3O+} at low cometocentric distances. As the cometocentric distance increases, the decrease in the coma density reduces the rate of collisions for chemical reactions to occur. Since the \ch{H2O+} density remains nearly constant, the decrease in the production of \ch{H3O+} is proportional to the decrease in the neutral density of its parent species (water). On the other hand, the decrease in the production of \ch{HPA+} (for example, \ch{NH4+}) is proportional to the decrease in the density of the parent HPA (\ch{NH3} in this example) as well as the decrease in the density of the ion from which it accepts \ch{H+} (for example, \ch{H3O+}). This causes \ch{H3O+} to be the dominant ion at the spacecraft location. The main contributing ions to PI$_{\text {other}}$ are \ch{O2+} and \ch{CH3OH+}, with a smaller contribution from \ch{CO2+}. At low distances, the density of PI$_{\text {other}}$ is higher than that of \ch{H2O+}. This is because \ch{H2O+} is destroyed rapidly due to collisions with \ch{H2O} while \ch{O2+} and \ch{CH3OH+} do not react with \ch{H2O}. Although \ch{CO} has a relative abundance higher than most other trace volatiles, it reacts with \ch{H2O} via ion-neutral and charge exchange reactions, and it does not contribute significantly to PI$_{\text {other}}$.

\begin{table}
	\caption{Parameters used to fit a power law to the total ion density for the lower and upper bounds.}	
	\label{table:fit}
	\setlength{\tabcolsep}{12pt}
	\begin{tabular}{ ccccc} 
		\hline
		\multirow{2}{*}{Molecule}                  &
		\multicolumn{2}{c}{Lower bound} &
		\multicolumn{2}{c}{Upper bound}    \\ \cline{2-5}
		& $c_1$         & $c_2$             & $c_1$ &$c_2$\\ 
		\hline 
		$I_{7}$ & -1.015 & 1.345e+03 & -1.012 & 1.638e+03 \\
		$I_{8}$ & -1.017 & 1.344e+03 & -1.014 & 1.814e+03 \\
		$I_{9}$ & -1.014 & 1.772e+03 & -1.013 & 2.059e+03 \\
		$I_{10}$ & -1.015 & 1.827e+03 & -1.012 & 2.745e+03 \\
		$I_{11}$ & -1.011 & 2.212e+03 & -1.010 & 3.791e+03 \\
		$I_{12}$ & -1.012 & 2.905e+03 & -1.008 & 4.506e+03 \\
		$I_{13}$ & -1.005 & 5.761e+03 & -1.003 & 1.078e+04 \\
		$I_{14}$ & -1.003 & 7.800e+03 & -0.995 & 1.764e+04 \\
		$I_{15}$ & -1.007 & 6.492e+03 & -0.994 & 1.949e+04 \\
		$I_{16}$ & -0.999 & 2.160e+04 & -0.995 & 2.614e+04 \\
		$I_{17}$ & -1.002 & 3.292e+04 & -0.987 & 3.979e+04 \\
		$I_{18}$ & -0.965 & 6.584e+04 & -0.961 & 6.962e+04 \\
		$I_{19}$ & -0.975 & 6.268e+04 & -0.968 & 7.194e+04 \\
		$I_{20}$ & -0.982 & 5.592e+04 & -0.966 & 7.202e+04 \\
		$I_{21}$ & -0.969 & 8.977e+04 & -0.937 & 1.344e+05 \\
		$I_{22}$ & -0.925 & 1.484e+05 & -0.919 & 1.653e+05 \\
		$I_{23}$ & -0.896 & 1.916e+05 & -0.882 & 2.063e+05 \\
		$I_{24}$ & -0.878 & 2.158e+05 & -0.856 & 2.149e+05 \\
		$I_{25}$ & -0.853 & 2.191e+05 & -0.840 & 2.160e+05 \\
		$I_{26}$ & -0.952 & 9.797e+04 & -0.879 & 1.934e+05 \\
		$I_{27}$ & -0.975 & 6.940e+04 & -0.930 & 1.306e+05 \\
		$I_{28}$ & -0.932 & 1.283e+05 & -0.920 & 1.433e+05 \\
		$I_{29}$ & -0.944 & 9.363e+04 & -0.940 & 1.055e+05 \\
		$I_{30}$ & -0.960 & 6.695e+04 & -0.949 & 8.743e+04 \\
		$I_{31}$ & -0.980 & 4.518e+04 & -0.972 & 5.506e+04 \\
		$I_{32}$ & -0.984 & 3.594e+04 & -0.980 & 4.397e+04 \\
		$I_{33}$ & -0.988 & 3.015e+04 & -0.987 & 3.420e+04 \\
		$I_{34}$ & -0.996 & 2.100e+04 & -0.995 & 2.631e+04 \\
		$I_{35}$ & -0.998 & 1.590e+04 & -0.998 & 1.881e+04 \\
		$I_{36}$ & -0.993 & 1.712e+04 & -0.992 & 1.973e+04 \\
		$I_{37}$ & -1.004 & 9.549e+03 & -1.000 & 1.490e+04 \\
		$I_{38}$ & -1.004 & 7.601e+03 & -1.006 & 1.007e+04 \\
		$I_{39}$ & -1.010 & 7.398e+03 & -1.008 & 8.888e+03 \\
		$I_{40}$ & -1.015 & 5.069e+03 & -1.013 & 6.302e+03 \\
		\hline
		
	\end{tabular}
\end{table}

In the first half of the group 3 time intervals (Figure \ref{fig:n3}), the ion composition shows similarities with the group 2 intervals, with \ch{H3O+} dominating at the spacecraft location, while at low cometocentric distances, \ch{HPA+} and PI$_{\text {other}}$ show densities that are higher than \ch{H3O+} and \ch{H2O+}, respectively. In fact, this is the trend followed in the intervals in which $Q>10^{27}$ s$^{-1}$. In the latter half, the density of \ch{HPA+} ions at low cometocentric distances is nearly equal to \ch{H3O+}, which is different from what we see for the pre-perihelion case of the group 1 intervals. This can be explained by the sharper decline in the \ch{H2O} sublimation rate post-perihelion, which results in a higher relative abundance of the HPA neutrals and higher density of \ch{HPA+}. The same is true for PI$_{\text {other}}$ ions that show a higher density than \ch{H2O+} for low cometocentric distances. In this case, \ch{CO2+} is the dominant ion contributing to PI$_{\text {other}}$ at the spacecraft location while most of the other ions show number densities of $\sim 0.5-3$ cm$^{-3}$. We have not shown the ion composition for the intervals $I_{41}$ and $I_{42}$ in Figure \ref{fig:n3}; for these intervals, the low spacecraft distance and the latitudinal variability lead to high uncertainties in modeling the coma chemistry.

The total ion densities shown in Figures $\ref{fig:n1} - \ref{fig:n3}$ follow a near power law radial dependence. We derived the power law of the total plasma number density $n_\text{{tot}}$ on the cometocentric distance $r$ as
	
\begin{equation}
n_\text{{tot}} = c_2 r^{c_1}.
\end{equation}
Table \ref{table:fit} shows the values of the fitted parameters $c_1$ and $c_2$ for the lower and upper bounds of the modeled number density in different time intervals. The ion density for most intervals varies approximately as $r^{-1}$ which is expected. However, for intervals near perihelion, when ion-neutral reaction rates are faster, there is a deviation from the $r^{-1}$ dependence.

\section{Discussions} \label{sec:dis}
\noindent Following the in situ detections of a large number of ions by \textit{Giotto} at 1P/Halley, \cite{Haider2005} showed that photochemical modeling can be used to estimate the densities of the detected ion peaks in the mass spectra. In anticipation of the \textit{Rosetta} observations near perihelion, \cite{Vigren2013} predicted the ion composition of 67P/C-G by assuming the neutral composition to be made up of \ch{H2O}, {CO} and 1\% of high proton affinity neutrals. \cite{Heritier2017b} improved the number density estimates for the near perihelion conditions using neutral measurements from ROSINA-COPS. In some studies, the ratios between the integrated ion counts of two ion species, such as \ch{NH4+}/\ch{H2O+}, \ch{NH4+}/\ch{H3O+} and \ch{H3O+}/\ch{H2O+} were used to estimate the ion densities in the coma \citep{Beth2016, Fuselier2015, Fuselier2016}. However, as pointed out by \cite{Heritier2017b}, a problem with this approach is that the DFMS scans for two ions are not simultaneous. Additionally, only a small part of the full ion distribution is sampled due to the narrow DFMS field of view and the inability of ions of all energies to pass through the DFMS electrostatic analyzer. 

From the results that we have presented in the previous section, we see that our photochemical model can provide an estimate towards the cometary ionospheric composition within factors of $1-3$ for the group 1 intervals and $2-3$ for the group 3 intervals. For the group 2 intervals, if we include UV extinction, the ion density is estimated within factors of $1-4$, though it is overestimated by up to a factor of 6 in the interval $I_{18}$. Additionally, if we consider the lower bound of the production rates provided by \cite{Lauter2020}, the modeled ion densities are closer to the observed densities, as compared to the upper bound. Our model runs show that for most of the time intervals in our study, \ch{H3O+} has the highest density at the location of \textit{Rosetta}, followed by \ch{H2O+}. The density of the \ch{HPA+} ion group at the spacecraft location is generally lower than that of \ch{H2O+}, though around perihelion, it becomes similar to or even larger than \ch{H2O+}. The ions belonging to the ion group PI$_{\text{other}}$ generally have the least abundance among all the ions considered. 

Since most cometary neutrals have ionization threshold energies of $12-14$ eV, \cite{Goetz2022} point out that the energy of photoelectrons created in the coma is not sufficient to cause electron impact ionization at high heliocentric distances ($>2$ au). At these distances, electron impact ionization is likely to be caused by solar wind electrons ($\sim 10$ eV) accelerated by the ambipolar electric field. In our models, at high heliocentric distances, the photoionization rates are $2-3$ orders of magnitude higher than the electron impact ionization frequency since we only consider impact ionization by photoelectrons. In order to model the impact ionization by solar wind electrons, we would have to include kinetic effects. This is currently beyond the scope of the current work though we do aim to incorporate these effects in the future. \cite{Galand2016} show that for some periods (i.e. when the spacecraft is in the northern hemisphere), photoionization alone can explain the ionospheric density, and the inclusion of electron impact ionization leads to an overestimation of the density. In the southern hemisphere, electron impact ionization is required to explain the ion density. Such seasonal variations are also noted by \cite{Heritier2018}.

Our assumption of a common bulk velocity for the ions and neutrals works well for low cometary activity and at a distance of several tens of kilometers from the nucleus, as also seen in other studies (for example, \citealt{Galand2016, Vigren2016, Heritier2017a, Heritier2018}). This assumption may not hold when the spacecraft is further away from the nucleus (beyond 100 km) and for the higher outgassing rates near perihelion. Ion velocity measurements near perihelion show values of $2-8$ km s$^{-1}$, as opposed to a neutral outflow velocity of $\sim 1$ km s$^{-1}$ \citep{Vigren2017}. This may be caused by the acceleration of the ions along the ambipolar electric field set up by the electron pressure gradient. \cite{Odelstad2018} find the ion velocity to be distributed around $3.5-4$ km s$^{-1}$ inside the diamagnetic cavity and higher velocities of $\lesssim8-10$ km s$^{-1}$ in the surrounding region, indicating that the ion-neutral drag force does not balance the outside magnetic pressure at the cavity boundary. These effects may result in an overestimation of the modeled density by factors of $2-5$ \citep{Vigren2019}.

The cometary ion flow patterns are also affected by the solar wind electric fields. Studies have identified two ion populations, namely the pick-up ion population that gains its energy and momentum through interactions with the solar wind upstream of the observation point and the expanding ion population of cometary origin that gains most of its energy in the vicinity of the nucleus \citep{Nilsson2015, Nilsson2017, Nilsson2020, Behar2016, Bercic2018}. While the motion of the former ion population is governed by the solar wind electric field, the latter ion population is accelerated by the ambipolar electric field in the vicinity of the nucleus and its motion is radially away from the comet nucleus in the $Y-Z$ plane perpendicular to the comet-Sun direction. This is in agreement with the strong shielding of the inner coma from the solar wind electric field at sufficiently high activity \cite{Nilsson2018}. Based on observations of anti-sunward streaming cometary ions, it is suggested that there is an anti-sunward component of the electric field that has a strength of about 10\% of the solar wind electric field. Although there is an asymmetry in the coma of 67P/C-G, there is not much influence in the flow direction \citep{Bercic2018}. Since most of the cometary ions are created by the ionization of neutrals that expand radially, \cite{Edberg2019} point out that it is reasonable to assume that the cometary plasma originates in the region between \textit{Rosetta} and the nucleus, as opposed to being ionized elsewhere in the coma and then transported to \textit{Rosetta} via another route.

Putting our results in the context of the spectra acquired by DFMS, we find that our results align with the DFMS observations. \ch{H3O+} and \ch{H2O+} were regularly detected during the entire mission period, with the largest counts observed near perihelion \citep{Fuselier2015, Fuselier2016,Beth2016, Beth2020, Heritier2017b}. DFMS scanned the range around 18 uq$^{-1}$ three times more often as compared to the other mass ranges, though \ch{NH4+} was mostly detected during the higher outgassing period of the comet around perihelion \citep{Beth2016, Beth2020}. The ions \ch{HCNH+}, \ch{H2COH+} and \ch{CH3OH2+} belonging to the \ch{HPA+} ion group are also detected near perihelion in the DFMS high resolution (HR) spectra though the detection frequency is lower than that of \ch{NH4+} \citep{Beth2020}. This is to be expected, since \ch{NH4+} is the most abundant ion in the \ch{HPA+} group, and 18 uq$^{-1}$ is also scanned more often. \cite{Lewis2023} suggest that in addition to protonation of \ch{NH3}, the dissociation of ammonium salts from cometary dust grains near DFMS may also be a source of \ch{NH4+}. However, there is no evidence for the detection of \ch{H3S+}, which is not understood. It may be related to the reduced energy acceptance of the instrument at higher uq$^{-1}$ \citep{Heritier2017b} or the suggested origin of \ch{H2S} from dust grains \citep{Calmonte2016}.

\cite{Beth2020} report on the detection of several of the ions belonging to the group PI$_{\text{other}}$ for mass-per-charge ratio $<40$ uq$^{-1}$. We find \ch{CO+} to be a primary contributor to PI$_{\text{other}}$ in the time intervals of group 1, and it was detected in HR at $r_h>2.2$ au pre-perihelion. \ch{CH3OH+} was detected in HR near perihelion, where we estimate it to have the most abundance, though there is no confirmed detection of \ch{O2+} at the same uq$^{-1}$.  \cite{Beth2020} find peaks in low resolution (LR) at 27 uq$^{-1}$ and 34 uq$^{-1}$ near perihelion and post-perihelion at $r_h>2.2$ au and these may be attributed to \ch{HCN+} and \ch{H2S+} respectively, though they were not detected in HR. There is a weak signal corresponding to \ch{CH4+} in HR at $r_h>2.2$ au post-perihelion. With regard to \ch{NH3+}, \cite{Beth2020} clearly detect it near perihelion with higher counts than \ch{CH3OH+}. Our models predict \ch{NH3+} to have a number density of a few cm$^{-3}$ at the spacecraft location, and not as high as \ch{CH3OH+}. We can explain this discrepancy in ion counts as follows. \ch{CH3OH+} has a mass-per-charge ratio of 32 uq$^{-1}$ that is nearly double that of the mass-per-charge ratio of \ch{NH3+} (17 uq$^{-1}$). There is a reduction in the energy acceptance window of DFMS as the mass-per-charge ratio increases \citep{Schlappi2011Thesis, Heritier2017b, Beth2020}. Additionally, near perihelion, the ions are accelerated through a negative spacecraft potential, which further reduces the maximum energy that an ion can have in order to be detected by DFMS. We refer the reader to \cite{Heritier2017b} for a detailed analysis of the instrument sensitivity towards a detectable range of energies.


\section{Conclusions} \label{sec:concl}
\noindent We have modeled the cometary ionosphere for 36 of the time intervals defined by \cite{Lauter2020}, by using their neutral outgassing rates as inputs in our gas phase coma model.
For most of the intervals, we find that our model-derived total ion densities can match the plasma densities measured by LAP/MIP within factors of $1-3$ for group 1, $1-4$ for group 2 (when UV extinction is included), and $2-3$ for group 3.
We further obtain the cometocentric distance variation of the density of the ions \ch{H2O+} and \ch{H3O+} and the ion groups \ch{HPA+} and PI$_{\text{other}}$. We find that \ch{H3O+} is the most dominant ion at the location of \textit{Rosetta}. In the intervals near perihelion, the \ch{HPA+} ion group is dominant at low cometocentric distances. The PI$_{\text{other}}$ ion group dominates over \ch{H2O+} at low cometocentric distances post-perihelion due to the rapid decline in the water production rate. 

The deviations in our modeled ion densities from the plasma density measured by LAP/MIP can result due to several reasons. The variation in the local plasma density measured by \textit{Rosetta} arises due to a host of factors which include day-night asymmetry, latitudinal variability, spacecraft maneuvers, and pickup by solar wind. Since the production rate within a particular time interval is fixed, the only variation that our model is able to capture is due to changes in the cometocentric distance location of the spacecraft. The production rates rely on the DFMS data for which \cite{Rubin2019b} estimate a 30\% uncertainty that may arise due to detector gain, sensitivity calibration, and fitting errors. Uncertainties in the production rates also arise due to the limited surface coverage which results in unknown emissions from not-seen surface elements. The photoemission current is seen to reduce near perihelion and this is attributed to the EUV extinction due to scattering by cometary nanograins \citep{Johansson2017}. Finally, the sublimation of volatiles from dust grains may also be an ion source in the coma, an effect which is not included in our model \citep{Calmonte2016, Lewis2023}.

\section*{Acknowledgements}
\noindent The computations were performed on the Param Vikram-1000 High Performance Computing Cluster of the Physical Research Laboratory (PRL), India. The work done at PRL is supported by the Department of Space, Government of India. The data on the solar spectral UV fluxes was accessed via the LASP Interactive Solar Irradiance Datacenter (\url{https://lasp.colorado.edu/lisird/}). The photo-absorption cross sections are available at the PHIDRATES database (\url{https://phidrates.space.swri.edu/}). The time series data on the global volatile production rates was accessed via \url{https://arxiv.org/abs/2006.01750}.
We express our gratitude to the ROSINA and RPC team, PI-ROSINA K. Altwegg (University of Bern, Switzerland), PI-LAP A. Eriksson (Swedish Institute of Space Physics, Uppsala, Sweden), and PI-MIP P. Henri (LPC2E/CNRS, Orl\'eans, France). We thank R. Gill, E. P. G. Johansson, and F. L. Johansson for the Rosetta RPC-LAP archive of derived plasma parameters. We are grateful to Kinsuk Acharyya and Shashikiran Ganesh of PRL for useful insights and discussions. We thank the anonymous reviewers for their comments which strengthened this manuscript.

\bibliographystyle{cas-model2-names}



\end{document}